\documentstyle[aps,multicol]{revtex}

\begin{document}

\title{Quantum Information is physical too...}
\author{Terry Rudolph\footnote{email: tez@physics.utoronto.ca}}
\address{Department of Physics, University of Toronto,\\ Toronto M5S 1A7,\\ Canada}  

\date{\today}

\maketitle

\begin{abstract}

We discuss the physical nature of quantum information, in particular focussing on tasks that are achievable by some physical realizations of qubits but not by others. 

\end{abstract}

\pacs{}   

\def\){\right)}
\def\({\left(}
\def\>{\rangle}
\def\<{\langle}
\def\be{\begin{equation}}
\def\ee{\end{equation}}
\def\bea{\begin{eqnarray}}
\def\eea{\end{eqnarray}}
\def\ba{\begin{array}}
\def\ea{\end{array}}
\def\nn{\nonumber}

\def\pra#1#2#3{Phys. Rev. A {\bf#1}, #2 (#3)}
\def\prl#1#2#3{Phys. Rev. Lett. {\bf#1}, #2 (#3)}
\def\jpb#1#2#3{J. Phys. B {\bf#1}, #2 (#3)}
\def\josab#1#2#3{J. Opt. Soc. Am. B {\bf#1}, #2 (#3)}
\def\optcom#1#2#3{Opt. Comm. {\bf#1}, #2 (#3)}

\begin{multicols}{2}

One of the most interesting developments of modern quantum mechanics has been the realization that quantum information is extremely useful stuff, despite the fact that to quantitatively define exactly what it is turns out to be a little tricky . This is in contrast to classical information theory, which has a very elegant and abstract formulation. From  classical information  theory we learn that the medium is irrelevant to the message, in the sense that our theory can be constructed without requiring a different formulation for every particular physical mechanism of transmitting the classical information. (This is not to deny, however, the necessity of broad considerations which can be traced back to the physical method of transmission, such as the error rate).

It is perhaps because of the high degree of abstractness attainable in classical information theory, that it was many years before it was realized that considering the physical nature of information carriers, in particular allowing them to be quantum particles obeying the laws of quantum mechanics, could lead to novel effects that lie outside the realm encompassed by classical information theory. Such new effects include quantum dense coding\cite{dense}, quantum teleportation\cite{tele} and  quantum computing\cite{qcomp} to name a few.  Perhaps the most striking example in the preceding list, from the perspective of defining quantum information,  is quantum teleportation. In teleportation a quantum state is transmitted in two completely seperate parts; the first part, which is transmitted instantaneously via a measurement incorporating one member of an entangled pair of particles, consists of purely quantum information; the second part is comprised of two bits of purely classical information. The instantaneous nature of quantum information transmission is one of several features that distinguish it from its classical counterpart. Another is the fact that it cannot be copied and hence amplified \cite{noclone}.

Despite the generality of classical information theory, it is {\it not} true that the theory captures all possible tasks that can be acheived by two parties who are allowed to exchange arbitrary classical physical objects. To see this consider a situation where the two communicating parties, Alice and Bob are seperated in space and do not share a common co-ordinate system (set of $x,y,z$ axes). If they are only allowed to communicate within the confines of classical information theory {\it i.e.} by transmisson of 1's and 0's, they will be unable to establish a set of parallel axes. One could imagine this classical communication as occurring through the exchange of black and white ping-pong balls say. On the other hand, if they were to exchange precisely oriented pencils, then the task would be easily achievable. Therefore if we want to characterize all the tasks achievable with classical resources, we must take into consideration the physical nature of those resources.

Of course the preceding example is not meant to be a profound insight into classical information theory. Rather it is to set us thinking along the track that if we are going to attempt to quantify quantum information as a resource then we should consider  the physical properties of the quantum information too. At the heart of almost all the new effects mentioned previously lies quantum entanglement. For this reason much attention has focussed on quantifying entanglement in a manner which reflects both its usefulness as a resource, as well as the ease of producing or transferring it \cite{entangmeas}.  These methods of quantifying entanglement have managed to attain an impressive degree of abstractness, and different measures are available depending on what particular problems you are interested in. In general these measures of entanglement do {\it not} depend on which particular type of physical systems are entangled. They depend only on the dimensions of the Hilbert space involved and the form of the state vector in a particular choice of factorization of the Hilbert space.  Thus a pair of photons with maximally entangled polarizations {\it i.e.} in the state 
\be
|\psi\>=\frac{1}{\sqrt{2}}\(|H\>|V\>-|V\>|H\>\),
\ee
is considered equivalent (from the perspective of current measures of entanglement) to a pair of entangled spin-$1/2$ particles in the state
\be
|\phi\>=\frac{1}{\sqrt{2}}\(|\uparrow_z\>|\downarrow_z\>-|\downarrow_z\>|\uparrow_z\>\).
\ee
Here $\{|H\>,|V\>\}$ refer to horizontally and vertically polarized photons, while $\{|\uparrow_z\>,|\downarrow_z\>\}$ refer to spin eigenstates in the $z$ direction. 

To see that the assumed equivalence is not always justified, imagine a situation where Charlie supplies Alice and Bob with $N$ maximally entangled spin-$1/2$ particles, as well as a classical communication channel, and instructs them to construct a set of parallel axes. It is not hard to see that they can accomplish the task to a high level of accuracy given enough entangled particles. Alice need only pick a preferred direction ${\bf A}_z$, and make measurements on a subset of her $N$ particles with a Stern-Gerlach device oriented in this direction. She then uses the classical communication channel to inform Bob as to the outcome of her measurements. He may then rotate his own Stern-Gerlach device until he finds an anti-correlation with her results (although in general there will be some more efficient joint measurement to determine the direction).  It must be emphasized that the information as to the special direction ${\bf A}_z$, is transmitted purely by the instantaneous  quantum information transfer, the classical channel carries no directional information.   The other directions may be transmitted using the remaining particles.

The ability to complete this task is not particularly mysterious, since it is a consequence of the fact that spin-$1/2$ particles are physical objects whose description requires some specification of directions in space. It also is not hard to see that such a task is not achievable by two parties who share $N$ maximally entangled scalar particles. Interestingly it also appears that $N$ photons in the state (1) do not suffice either, since photon polarizations are constrained to lie in a plane perpendicular to their direction of propagation. Thus unless we have special knowledge about the preparation of the photons (for example that they are emitted in opposite directions or $6^0$ apart), we cannot deduce a correlation between the measured polarizations and ``absolute'' directions in space.

At this stage a clarifying remark regarding the physical nature of different possible qubit realizations should probably be made. Obviously there exist some two-state quantum systems which have ``natural'' orthogonal bases, for example the ground and excited states of an atom, or the zero and one mode excitations of a scalar field. These are energy eigenstates. Other two-state systems however have no such universally preferred basis, spin being a case in point. The difference lies purely in the Hamiltonian of the universe - if the universe was permeated by a huge static magnetic field then there {\it would} be a ``natural'' basis for spin measurements! The distinction is not purely academic however. Quantum teleportation, for example, requires the receiver to perform one of four unitary transforms to a qubit in order to recover the teleported state. If the sender and receiver do not share knowledge of an orthogonal basis then this is not possible.

Finally we give a rather cute example of ``Quantum Navigation'', which applies the preceding ideas to a problem which one hopes will not be faced by the reader, but which is worth bearing in mind when visiting politically unstable countries.
Consider the following (rather contrived) scenario: Bob's friend Alice has been abducted. She regains consciousness to find herself locked in a windowless room, presumably somewhere on Earth. Checking her pockets she finds that she still has her cellular telephone, and so she calls Bob. After being reassured that she's unhurt, Bob naturally asks ``Where are you?''. Of course Alice has no idea, and they discover that the telephone company is singularly unhelpful and refuses to co-operate in Bob's attempt to locate Alice. (This part of course is the most realistic part of the story). Fortunately Alice realizes that she still has in her pocket the box of $N$ spin-$\frac{1}{2}$ particles that are entangled (in singlet states) with particles in Bob's possession. The question we are interested in is how may these particles be used to locate Alice? (In what follows we ignore any possible effects on the particle's spin from Alice's journey through the Earth's magnetic field.)

The simplest solution is the following: Alice alignes her mini Stern-Gerlach device (which will soon be standard on all Swiss-Army knives) in {\it her} vertical direction (${\bf A}_z$). She then proceeds to measure the spin of each of her particles in this direction, announcing to Bob over the phone what the result ($\uparrow,\downarrow$) of each measurement is. Bob is now left with $N$ particles, $n_+$ of which are in the state $|\uparrow_{{\bf A}_z}\>$, while the remaining $n_-=N-n_+$ are in the state $|\downarrow_{{\bf A}_z}\>$. In general we expect $n_+\approx N/2$. If Bob can determine the direction ${\bf A}_z$, then he can determine where on Earth Alice is, since each position on earth has a unique $z$-direction defined by gravity. Of course since they only share a finite number of particles he will only be able to do this to a certain accuracy, but hopefully it is good enough that he can locate her and save her from her nefarious imprisoners.

An interesting question this now raises is what measurement strategy should Bob follow in order to obtain the best estimate as to the direction ${\bf A}_z$? One strategy is to make  joint measurements on the $n_+$  and $n_-$ sets of  particles seperately (since it is well known that a joint measurement on all the particles in each set would be preferable for estimating their state \cite{derka}), and then to average the result somehow. However Gisin and Popescu recently showed in \cite{gisin}, that a joint measurement on a pair of anti-parallel spins is in fact more efficient than that on a pair of parallel ones, given the task of determining as accurately as possible the direction the spin is pointing! If the result generalizes to more than two particles, then in fact Bob would do better to apply a joint measurement to all $N$ particles. This of course is not so surprisng. What {\it is} really amazing however, is that he would do {\it better} than if Alice had been allowed to send him a carefully prepared box of spins all pointing in the same direction. The randomness of the outcomes of her measurements would actually be helping the task at hand! This appears to be a clear violation of the Law of Conservation of Trouble. (While no precise formulation of the Law exists in non-relativistic quantum mechanics, the quantum field theory version can be found in \cite{weinberg}, and of couse we are all familiar with the Law in the macroscopic limit, whence it is more often called Murphy's Law.)

In conclusion it is has been shown that the physical nature of quantum information should be taken into account when trying to quantify it as a resource. A task was presented which can be completed successfully by holders of maximally entangled spin-$1/2$ particles, which cannot be completed successfully by holders of some other types of maximally entangled particles. Finally it was shown how these ideas could be used to allow an abducted person to communicate their position on earth, even though they have no knowledge of where they are themselves.
 
\acknowledgements
The author wishes to acknowledge many useful discussions and several stimulating arguments with R. Spekkens, as well as useful discussions with both A. Steinberg and J. Sipe.

\end{multicols}

\end{document}